\documentclass[prl,showpacs,twocolumn]{revtex4}

\usepackage{graphicx}
\usepackage{bm}

\newcommand{\bfr}{{\bf r}}
\newcommand{\nhat}{{\bf \hat{n}}}

\begin{document}

\title{Organization and instabilities of entangled active polar filaments.}
\author{Tanniemola B. Liverpool$^{1,3}$ and M. Cristina Marchetti$^{2,3}$}
\affiliation{$^1$Condensed Matter Theory Group, Blackett Laboratory, 
Imperial College, London, SW7 2BW\\
$^2$Physics Department, Syracuse University, Syracuse, NY 13244\\
$^3$Kavli Institute of Theoretical Physics, University 
of California, Santa Barbara, CA 93106}

\date{\today}

\begin{abstract}
We study the dynamics of an entangled, isotropic
solution of polar filaments coupled by molecular motors 
which generate relative motion of the filaments in two and three
dimensions. We investigate the stability of the homogeneous
state for constant motor concentration 
taking into account  excluded volume and
entanglement. At low 
filament density the system develops a {\em density} instability, while at
high filament density entanglement effects 
drive the instability of
{\em orientational} fluctuations.

\end{abstract}
\maketitle

Cellular biology provides many realizations of pattern formation in
dissipative nonequilibrium systems.  An important example is the
collective behavior of the proteins that compose the cytoskeleton of
eukaryotic cells. The cytoskeleton provides both the supporting
structure of the cell and the vehicle for internal transport processes
\cite{howard}. It is a network of long protein filaments, mainly
microtubules, actin filaments and intermediate filaments, coupled by
smaller proteins, such as molecular motors and cross-linkers.  Motor
proteins 
convert chemical energy derived
from the hydrolysis of ATP (Adenosine TriPhosphate) into mechanical
work, generating forces and motion of the filaments relative to each
other in these {\em active gels}.

Numerous {\em in vitro}
experiments~\cite{nedelec97,surrey01,humphrey02,loic} have shown that
mixtures of filaments and their associated motor proteins
self-organize into 
macroscopic symmetry-breaking
structures, including radial arrays or asters and one-dimensional
bundles.  The nonequilibrium forces that give rise to these structures
include the action of molecular motors and the
polymerization/depolymerization process of the filaments
\cite{biochem}.  Here we focus on the role of motor proteins and
assume that the filaments have fixed length {---} a situation that can
be achieved in vitro \cite{nedelec97}.  A few analytical and numerical
studies have investigated the emergence of these complex 
patterns~\cite{nakazawa96,nedelec97,bassetti00,kruse00,surrey01,lee01,kruse01}.
Continuum models of filament/motor systems in two dimensions have been
used to show that spatial patterns are obtained as
nonequilibrium solutions of the system
dynamics~\cite{lee01,bassetti00}.
These models have ignored
either filament diffusion \cite{lee01} 
or the motor action on orientational dynamics \cite{bassetti00}. 
A more microscopic
approach was taken by Kruse {\em et al} who considered a
dynamical model for the development of contractile and motile
structures in {\em one dimensional} polar filament bundles, while
ignoring steric and other interactions between the
filaments~\cite{kruse00,kruse01}.

Many open questions remain concerning the 
role of the physical properties of the filament/motor gel
in controlling the formation of self-organized structures.
Experiments have indicated that 
motor properties, such as their processivity 
{--} the fraction of time 
in a cycle a motor remains attached to the filament,
strongly influence pattern formation.
This is evident by comparing {\em in vitro} experiments
in microtubules-kinesin to those in  actin-myosin mixtures.
At high
motor concentration, microtubule-kinesin mixtures 
readily organize in a variety of spatial patterns,
provided the motors stall
at the polymer ends before detaching \cite{nedelec97,surrey01}.
In contrast, the homogeneous state is much more robust in 
the weakly coupled actin-myosin II systems,
where spatially inhomogeneous
structures develop only upon depletion of ATP 
or at much higher filament concentration \cite{kas02}.  
The physical characteristics of the filaments,
such as their persistence length, may also contribute
to the different behavior of these two active gels, as
actin-myosin networks are more
strongly entangled than microtubule-kinesin mixtures. Both
analytic theories~\cite{kruse00,kruse01,lee01} and
simulations~\cite{nedelec97,surrey01} have so far entirely
{\em neglected} entanglement.

In this letter we start from a phenomenological model 
in the spirit of Kruse et al. \cite{kruse00,kruse01} and obtain a
set of continuum equations to describe the dynamics and organization
of polar filaments driven by molecular motors in an unconfined
geometry in (quasi-)two and three dimensions ($d=2,3$). 
By modeling the motor-filament interaction microscopically,
we can calculate the magnitude and, most
importantly, the sign of the parameters of the continuum equations,
which cannot be obtained by symmetry arguments. We consider a
isotropic filament solution and include the effects of {\em
  entanglement} and {\em excluded volume}. Our result
is a phase diagram (Fig.~\ref{fig:phasediagram}) as a function of the
filament density and motor properties.

The filaments are modeled as rigid rods of length
$l$ and diameter $b<<l$. Each filament is identified by the
position $\bfr$ of its center of mass and a unit vector $\nhat$
pointing towards the polar end.  Taking into account {\em filament
  transport}, the normalized filament probability distribution
function, $\Psi(\bfr,\nhat,t)$, obeys a conservation
law~\cite{DoiEdwards},
\begin{equation}
\partial_t \Psi +\nabla\cdot{\bf J}+\bf{\cal R}\cdot{\bf J^r}=0\;,
\label{conservation}
\end{equation}
where ${\cal R}=\nhat\times\partial_{\nhat}$ is the rotation operator.
The translational and rotational currents ${\bf J}$ and ${\bf J^r}$
are given by
\begin{eqnarray}
\label{Jt}
& & J_i=-D_{ij}\partial_j \Psi
         -\frac{D_{ij}}{k_BT}~\Psi~\partial_jV_{\rm ex}
         +J^{\rm act}_i\;,\\
\label{Jr}
& & J_i^r=-D_r{\cal R}_i \Psi
        -\frac{D_r}{k_BT}~ \Psi ~{\cal R}_iV_{\rm ex}
        +J^{\rm r/act}_i\;,
\end{eqnarray}
where $i=1,...,d$ and $D_{ij}=D_\parallel\hat{n}_i\hat{n}_j+D_\perp(\delta_{ij}
-\hat{n}_i\hat{n}_j)$ is the translational diffusion tensor and $D_r$ the
rotational diffusion constant.
The potential $V_{\rm ex}$ incorporates excluded volume effects that
play an important role in stabilizing time-dependent solutions. 
It is given by $k_BT$ times the probability of finding another rod in the
interaction area of a given rod,
\begin{equation}
\label{Vex}
V_{\rm ex}(\bfr,\nhat_1)=k_BT\int_{\nhat_2}\int'_{\bm{\xi}}
           \Psi(\bfr+\bm{\xi},\nhat_2)\;,
\end{equation}
where the prime restricts the integral to the interaction volume,
corresponding to the region where the 
two filaments touch at at least one point.
The volume of this region is 
$V_{\rm int}=v_0\sqrt{1-(\nhat_1\cdot\nhat_2)^2}$,
with $v_0=l^2 b^{d-2}$ and $l^2\sqrt{1-(\nhat_1\cdot\nhat_2)^2}>b^2$.
The active currents 
are given by
\begin{eqnarray}
\label{Jact}
& &\hspace{-0.2in}\!\!{\bf J}^{\rm act}(\bfr,\nhat_1)=\!\!\int_{\nhat_2}\!\int'_{\bm{\xi}}
         \!\! {\bf v}(\bm{\xi},\nhat_1,\nhat_2)
         \Psi(\bfr,\nhat_1)\Psi(\bfr+\bm{\xi},\nhat_2),\\
\label{Jr/act}
& &\hspace{-0.2in}\!\!{\bf J}^{\rm r/act}(\bfr,\nhat_1\!)\!=\!\!\!
       \int_{\nhat_2}\!\int'_{\bm{\xi}}
          \!\!\!\bm{\omega}(\bm{\xi},\nhat_1,\nhat_2)
       \Psi(\bfr,\nhat_1)\Psi(\bfr+\bm{\xi},\nhat_2),
\end{eqnarray}
\noindent where ${\bf v}=-\dot{\bm{\xi}}$  
and $\bm{\omega}=\dot{\nhat}_1-\dot{\nhat_2}$ are the relative linear
and angular velocities of two filaments, with the dot denoting a time
derivative.  The model naturally contains two competing dynamics. The
first is the  diffusion of hard rods, which at high
density must include excluded volume and entanglement. The
second is the local driving force coming from the interaction with the
motors. This depends on the polarity of the filaments and breaks the
$\nhat\rightarrow -\nhat$ symmetry of the hard rod fluid, allowing
for states of broken symmetry, where the filaments acquire a
nonvanishing mean orientation. 
\begin{figure}
\resizebox{3.truein}{!}
{\includegraphics{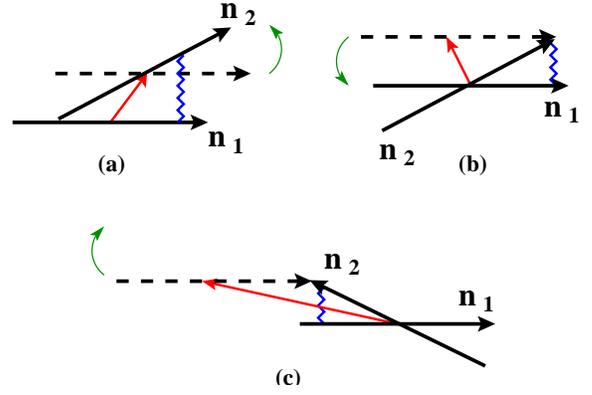}}
\caption{\label{fig:interaction} 
Cartoons of various motor-induced filament interactions.
All interactions are viewed from the rest frame of 
filament 2.
The initial and final position of filament 1 are shown as a 
thick and a dashed arrow, respectively.
The zig-zaged lines represent motors.
(a) The contribution to ${\bf v}$ proportional to
$\alpha$ which is along the direction of the relative
displacement $\bm{\xi}$ of the centers of mass of the two filaments
(thin arrow). Motors also drive 
rotation of filament 1 as indicated.
The contribution to ${\bf v}$ proportional to
$\beta$ is illustrated in
(b) and (c) for
two filaments with $\bm{\xi}=0$ and $\nhat_1\cdot\nhat_2>0$
(b) and $\nhat_1\cdot\nhat_2<0$ (c). In both cases the translation
at a rate $\beta$
in the direction 
of $\nhat_2-\nhat_1$ (thin arrow) tends to bring the
polar heads of the two filaments to the same spatial location. 
In (b) the counterclockwise rotation aligns 
the filaments, while in (c) the clockwise  rotation
anti-aligns and 
separates.
}
\end{figure}

In the absence of external forces and torques 
the total linear and angular velocity of an interacting pair are
conserved. This requires ${\bf v}(\bm{\xi},\nhat_1,\nhat_2)=-{\bf
v}(-\bm{\xi},\nhat_1,\nhat_2)$ and
$\bm{\omega}(\bm{\xi},\nhat_1,\nhat_2)=-\bm{\omega}(-\bm{\xi},
\nhat_1,\nhat_2)$. Rotational
and translational invariance requires ${\bf
v}(\bm{\xi},\nhat_1,\nhat_2)=-{\bf v}(-\bm{\xi},-\nhat_1,-\nhat_2)$
and $\bm{\omega}(\bm{\xi},\nhat_1,\nhat_2)=\bm{\omega}
(-\bm{\xi},-\nhat_1,-\nhat_2)$.  The simplest form of the linear and
angular velocities can be written as
\begin{eqnarray}
\label{v}
& & {\bf v}=\frac{\alpha}{2l}~\frac{\bm{\xi}(1+\nhat_1\cdot\nhat_2)}
        {\sqrt{1-(\nhat_1\cdot\nhat_2)^2}}
        +\frac{\beta}{2}\frac{\nhat_2-\nhat_1}
        {\sqrt{1-(\nhat_1\cdot\nhat_2)^2}}\;,\\
\label{omega}
& & \bm{\omega}=\gamma~(\nhat_1\cdot\nhat_2)~\frac{\nhat_1\times\nhat_2}
        {\sqrt{1-(\nhat_1\cdot\nhat_2)^2}}\;.
\end{eqnarray}
The velocities have been normalized with the volume of interaction. 
The parameters $\alpha$, $\beta$ and $\gamma$ are the rates for the various
motor-induced translations and rotations (Fig.~\ref{fig:interaction}). 
The contribution proportional to $\alpha$ depends on the separation of
the centers of the filaments and results from a difference in motor
activity between the ends and mid-points of the filaments. It tends to
align the centers of mass and polar heads of the filament pair (see Fig
2(a)).
The contribution
proportional to $\beta$ vanishes for aligned filaments and can separate
antiparallel filaments, as illustrated in Fig. 2(c). This mechanism 
yields both 
translational and rotational currents. The prefactor 
$(\nhat_1\cdot\nhat_2)$ in the angular velocity guarantees that 
motors preferentially bind to two filaments that are at an angle smaller
than $\pi/2$.
The $\gamma$ term has no effect on perpendicular filaments.
Assuming uniform motor density $\rho_m$, from simple mechanical models
of motors~\cite{howard}, we estimate $\alpha\simeq\beta\simeq {\gamma
  l} \simeq \rho_m l b^2\phi (s_c/\tau_c)$, with $s_c$
the motor step length per cycle, $\tau_c$ the time for one cycle and
$\phi$ the duty ratio.

We are interested in describing the dynamics of active 
filaments on length scales large compared to the filaments size, $l$.
We can then 
expand the concentration of filaments  $\Psi(\bfr+\bm{\xi},\nhat_2)$ 
near its value at $\bfr$, 
\begin{eqnarray}
\label{expansion}
\Psi(\bfr+\bm{\xi},\nhat_2) =&& \Psi(\bfr,\nhat_2)
  +\xi_n{\bf \hat{e}}_n\cdot\nabla \Psi(\bfr,\nhat_2)\nonumber\\
 & &\hspace{-0.8in}+\frac{1}{2}\xi_n\xi_m({\bf \hat{e}}_n\cdot\nabla)
         ({\bf \hat{e}}_m\cdot\nabla)\Psi(\bfr,\nhat_2)+O(\xi^3).
\end{eqnarray}
We have introduced  a set of orthogonal
unit vectors,  $({\bf \hat{e}}_1,{\bf \hat{e}}_2,{\bf \hat{z}})$,
that provides a natural coordinate system for the problem.
The unit vector ${\bf \hat{z}}$ is
normal to the plane passing through the point of contact
of the two filaments and containing the unit vectors
$\nhat_1$ and $\nhat_2$. The vectors
${\bf \hat{e}}_1=(\nhat_1+\nhat_2)/|\nhat_1+\nhat_2|$ and 
${\bf \hat{e}}_2={\rm sign}\big(\nhat_1\cdot\nhat_2\big)(\nhat_2-\nhat_1)/|\nhat_2-\nhat_1|$ are orthogonal unit vectors in this plane.
Neglecting the out-of-plane separation (of order $b$) between the
centers of mass of the two filaments, the vector $\bm{\xi}$
 is written in this coordinate system 
as $\bm{\xi}=\xi_n{\bf \hat{e}}_n$,
where summation over $n=1,2$ is intended.
We assume that on large scales the filament dynamics 
can be described in terms of the filaments density $\rho(\bfr)$ and 
the local filament orientation ${\bf t}(\bfr)$ defined as the first 
two moments of the distribution $\Psi(\bfr,\nhat,t)$,
\begin{equation}
\label{rho&t}
{\rho(\bfr,t) \choose {\bf t}(\bfr,t)}
   =\int d\nhat ~{1 \choose \nhat}\Psi(\bfr,\nhat,t)\;.
\end{equation}
Coarse-grained equations for $\rho$ and ${\bf t}$ can be obtained by
inserting Eq.~(\ref{expansion}) in the expressions for the active
currents and for $V_{\rm ex}$, writing the density
$\Psi(\bfr,\nhat,t)$ in the form of an exact moment expansion, and
retaining only the first two moments in this expansion.  For brevity,
we only display here the dynamical equations linearized about a
homogeneous state, with constant density $\rho_0$ and an isotropic
orientational distribution of filaments, corresponding to ${\bf t}=0$.
The full and rather cumbersome nonlinear equations will be given
elsewhere \cite{web_ref}.  
Letting $\rho=\rho_0+\delta\rho$ and keeping only terms up
to third order in the gradients, the linearized equations are
given by
\begin{widetext}
\begin{eqnarray}
\partial_t\delta\rho=\frac{1}{d}\big[D_\parallel+(d-1)D_\perp\big]
        (1+v_0\rho_0)\nabla^2\delta\rho
   -\frac{\alpha lv_0\rho_0}{12d}\nabla^2\delta\rho
   -\frac{\beta l^2v_0\rho_0(2d+1)}{24d(d+2)}\nabla^2(\bm{\nabla}\cdot{\bf t})\;,
\end{eqnarray}
\begin{eqnarray}
\label{orientation}
\partial_t t_i&=& -D_rt_i
   +\frac{1}{d+2}\big[(d+1)D_\perp+D_\parallel\big]\nabla^2t_i
           +\frac{2}{d+2}\big(D_\parallel-D_\perp\big)
           \partial_i\nabla\cdot{\bf t}\nonumber\\
& & -\frac{\alpha lv_0\rho_0}{12d(d+2)}
   \big[\nabla^2t_i+2\partial_i\bm{\nabla}\cdot{\bf t}\big]
   +\frac{\beta v_0\rho_0}{d}\partial_i\delta\rho
   +\frac{\beta l^2v_0\rho_0(2d+1)}{24d^2(d+2)}\partial_i\nabla^2\delta\rho
 \;.
\end{eqnarray}
\end{widetext}
The local orientation is not a conserved variable and decays
at a rate $\sim D_r$. Both equations 
display the competition of diffusive terms ($\propto D\nabla^2$)
and pattern-forming terms ($\propto -\alpha\nabla^2$). 
The linear instability of the
homogeneous state occurs when the pattern-forming terms dominate.
To linear order, the contribution from the rotational current
vanishes and excluded volume corrections only appear in the density equation.

We now study the stability of the homogeneous state. 
We expand the fields in Fourier components,
%
%
$\delta\rho(\bfr)=\sum_{\bf k}\rho_{\bf k}e^{i{\bf k}\cdot\bfr}$
and ${\bf t}(\bfr)=\sum_{\bf k}{\bf t}_{\bf k}e^{i{\bf k}\cdot\bfr}$,
and separate ${\bf t}_{\bf k}$ in its component longitudinal and transverse to ${\bf k}$, namely 
$t^L_{\bf k}={\bf \hat{k}}\cdot{\bf t}_{\bf k}$
and $t^T_{\bf k}={\bf \hat{k}}\times{\bf t}_{\bf k}$,
with ${\bf \hat{k}}={\bf k}/|{\bf k}|$. In $d$ dimensions there are 
$d-1$ degenerate transverse modes describing the decay of fluctuations
in $t^T_{\bf k}$, with rate
\begin{equation}
\lambda_T(k)=-D_r-\frac{k^2}{d+2}\Big[(d+1)D_\perp+D_\parallel-
    \frac{\alpha lv_0\rho_0}{12d}\Big]\;.
\end{equation}
There are two coupled modes describing the decay of density and $t^L_{\bf k}$
fluctuations, given by
\begin{equation}
\lambda_\pm(k)=\frac{1}{2}\Big\{M_{11}+M_{22}\pm
   \sqrt{(M_{11}-M_{22})^2+4M_{12}M_{21}}\Big\}\;,
\end{equation}
with 
\begin{eqnarray}
& & M_{11}=-\frac{k^2}{d}\Big[(D_\parallel+(d-1)D_\perp)(1+v_0\rho_0)
      -\frac{\alpha lv_0\rho_0}{12}\Big]\;,\nonumber\\
& & M_{22}=-D_r-\frac{k^2}{d+2}\Big[3D_\parallel+(d-1)D_\perp
      -\frac{\alpha lv_0\rho_0}{4d}\Big]\;,\nonumber\\
& & M_{12}=ik^3\frac{\beta l^2v_0\rho_0}{24}\frac{2d+1}{d(d+2)}\;,\nonumber\\
& & M_{21}=ik\frac{\beta v_0\rho_0}{d}\Big(1-\frac{l^2k^2}{24}\frac{2d+1}{d(d+2)}\Big)\;.
\end{eqnarray}
At long wavelength, $\lambda_+ \equiv \lambda_\rho$ vanishes as $k^2$
and describes the decay of density fluctuations, while $\lambda_-
\equiv \lambda_L$ is a kinetic mode (i.e., it is finite as
$k\rightarrow 0$) and describes the decay of longitudinal orientation
fluctuations.  The hydrodynamic density mode goes unstable on all
length scales for $\tilde{\alpha}>\tilde{\alpha}_c^{(1)}
\simeq(1+(d-1)D_\perp/D_\parallel)[1+(\tilde{\rho}_0)^{-1}]$, where
$\tilde{\alpha}=\alpha l/(12 D_\parallel)$ and
$\tilde{\rho}_0=\rho_0v_0$.

\paragraph{Dilute Solutions.}
For dilute solutions of long thin rods the diffusion constants are
$D_\perp=D_\parallel/2=D/2$ and $D_r=6D/l^2$, with
$D=k_BT\ln(l/b)/(2\pi\eta l)$ and $\eta$ the solvent viscosity
\cite{DoiEdwards}. In this regime the
instability of the homogeneous solution occurs at
$\tilde{\alpha}_c^{(1)}$ and is associated with the density
mode. 

\paragraph{Semi-dilute Solutions.} 
For semidilute solutions, the dynamics is modified by the topological
constraint that the filaments cannot pass through each other. This
constraint can be modeled by a tube~\cite{DoiEdwards,tube} leading to
modified transverse and rotational diffusivities, $D_\perp\simeq
D/[2(1 + c_\perp \tilde{\rho}_0 (l/b)^{d-2})^{2}]$ and $D_r \simeq
6 D/[l^2 (1+ c_r \tilde{\rho}_0 (l/b)^{d-2})^{2}]$ with $c_{r,\perp}$
numbers of order unity  and $D_\parallel$
essentially unaffected by entanglement \cite{DoiEdwards,tube}. The
kinetic modes $\lambda_-$ and $\lambda_T$ describing the decay of
orientation fluctuations can then become unstable before the density
mode. This instability occurs at
$\tilde{\alpha}_c^{(2)}\approx 1/\tilde{\rho}_0$ for
$k>k_0$, with $k_0\simeq(1/l\tilde{\rho}_0)
\sqrt{\frac{d(d+2)}{3(\tilde{\alpha}\tilde{\rho}_0-d)}}$.  A stability
diagram in the $(\tilde{\alpha},\tilde{\rho}_0)$ plane is shown in
Fig.~\ref{fig:phasediagram}. 
%

\paragraph{Semiflexibility.} Semiflexible filaments of persistence length 
$\ell_p > l$ can be modeled as fuzzy rods with a diameter given by $b
\equiv \sqrt{l^3/\ell_p}$ leading to $v_0 = l^{2 + 3(d-2)/2}
\ell_p^{-(d-2)/2}$.

\begin{figure}
\resizebox{3.truein}{!}
{\includegraphics{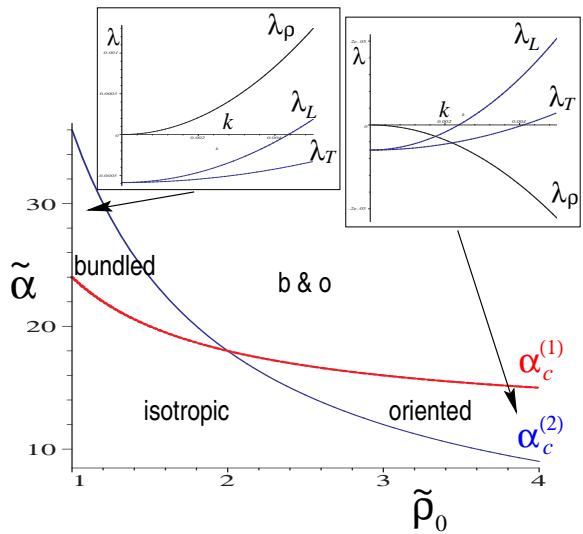}}
\caption{\label{fig:phasediagram} 
  Linear modes and phase diagram $\{\tilde{\alpha},\tilde{\rho_0}\}$
  showing the phase boundaries $\alpha_c^{(1)},\alpha_c^{(2)}$ between
  the homogeneous state and the density and orientationally
  inhomogeneous state.  The modes on the right corresponds to $\alpha l
  /D = 10, \tilde{\rho}_0=10, l/b =100$ while those on the left
  $\alpha l /D = 1900, \tilde{\rho}_0=1, l/b =100$.
}
\end{figure}

\paragraph{Discussion.}
The linear instability of the homogeneous state is controlled by the
parameter $\alpha$ that drives filament bunching, while $\beta$ and
$\gamma$ play no role \cite{other_param}.  At low filament densities,
well below the critical density for the nematic transition, density
fluctuations become unstable on all length scales at $\alpha_c^{(1)}$,
signaling the onset of a state with inhomogeneous density.  For
$\tilde{\rho}_0>\tilde{\rho}_c\simeq 2$, where the critical curves
$\alpha_c^{(1)}$ and $\alpha_c^{(2)}$ cross, the instability occurs at
the lower value $\alpha_c^{(2)}$ and it corresponds to a short-scale
instability of orientation fluctuations, while the density remains
homogeneous. In this regime the filament solution has a substantial
degree of short range nematic order and both transverse and rotational
diffusion are strongly impeded. The orientational degree of freedom,
${\bf t}$, becomes essentially conserved as $D_r$ is vanishingly
small.  The mechanism for the orientational instability of dense
filament solutions obtained here is distinct from that proposed by Lee
and Kardar \cite{lee01}. These authors modeled the dynamic in the
nematic phase and incorporated an inhomogeneous motor density. In
their model the instability is controlled by the nonlinear coupling of
orientation to motor density fluctuations. Here, in contrast the
instability occurs to linear order and is driven by the
pattern-forming terms in Eq.~(\ref{orientation}).

A number of open questions remain.  It will be 
relatively straightforward to include the motor transport
using our formalism.  This is important for very processive motors
and at low motor densities.  An inhomogeneous motor density may also be
required for the formation of stable asters and vortices at low
filament concentration.  It will be also very interesting to ask about the
response of the motor/filament system to shear.
This work will be presented elsewhere~\cite{unpublished}.

We thank E. Frey, K. Kruse, F. MacKintosh, A.C. Maggs, D. Morse and S.
Ramaswamy for many helpful discussions.  This research was supported
in part by the National Science Foundation under Grants No.
PHY99-07949 (at KITP), DMR97-30678 (at Syracuse) and the Royal
Society (London).
%


\begin{thebibliography}{99}
\bibitem{biochem}
Pattern formation in cells is of course regulated by 
complex biochemical reaction networks.
Here we focus only on the underlying physical processes.
\bibitem{other_param}
$\beta$ controls, however, the length scale at which
the modes are stabilized at higher wavevectors.
\end{thebibliography}

\end{document}